\documentclass[bm,aps,showpacs,amsfonts,amssymb,preprint,nofootinbib]{revtex4}

\usepackage{graphicx}
\usepackage{natbib}
\usepackage{amsmath}

\usepackage{color}
\font\mybb=msbm10 at 12pt

\font\mybbsub=msbm10 at 8pt

\def\bb#1{\hbox{\mybb#1}}

\def\bbsub#1{\hbox{\mybbsub#1}}

\def\RR {\bb{R}}

\def\RRsub {\bbsub{R}}

\def\CC {\bb{C}}
\def\CCsub {\bbsub{C}}
\def\HH {\bb{H}}
\def\HHsub {\bbsub{H}}
\def\AA {\bb{A}}
\def\AAsub {\bbsub{A}}

\def\OO {\bb{O}}
\def\OOsub {\bbsub{O}}

\newcommand\beqa{\begin{eqnarray}}
\newcommand\eeqa{\end{eqnarray}}
\newcommand\n{\nonumber\\}
\begin{document}

{~}

\title{On Dimensional Reduction of Magical Supergravity Theories}
\vspace{2cm}
\author{Naoto Kan\footnote[1]{E-mail:naotok@post.kek.jp} and Shun'ya Mizoguchi\footnote[2]{E-mail:mizoguch@post.kek.jp}}
\vspace{2cm}
\affiliation{\footnotemark[1]\footnotemark[2]SOKENDAI (The Graduate University for Advanced Studies)\\
Tsukuba, Ibaraki, 305-0801, Japan 
}
\affiliation{\footnotemark[2]Theory Center, Institute of Particle and Nuclear Studies,
KEK\\ Tsukuba, Ibaraki, 305-0801, Japan 
}
\begin{abstract} 
We prove, by a direct dimensional reduction and an explicit  
construction of the group manifold, that the nonlinear sigma 
model of the dimensionally reduced three-dimensional 
$\AA=\RR$ magical supergravity is 
$F_{4(+4)}/(USp(6)\times SU(2))$. 
This serves as a basis for the solution generating technique in this 
supergravity as well as allows to give the Lie algebraic characterizations 
to some of the parameters and functions in the original 
$D=5$ Lagrangian. Generalizations to other magical supergravities 
are also discussed.

\end{abstract}

\preprint{KEK-TH 1901}
\date{May 6, 2016}

\maketitle

\section{Introduction}
One of the common features of dimensionally reduced supergravity 
theories is that they contain a noncompact scalar coset sigma model 
in their Lagrangians. Perhaps 
the most famous example is the $E_{7(+7)}/SU(8)$ 
coset in $D=4$ ${\cal N}=8$ supergravity \cite{CJ} obtained by a 
dimensional reduction of the eleven-dimensionally supergravity 
\cite{CJS} to four dimensions. It is always the case that the global 
symmetry of the nonlinear sigma model is a symmetry of 
the whole supergravity system including the fermionic sector. 
A reduction of the eleven-dimensional supergravity 
to an intermediate dimension from 5 to 10
also yields an $E$-series symmetry \cite{JuliaGroupDisintegrations,Keurentyes},
whose discrete subgroup is nowadays understood as a 
U-duality \cite{HullTownsend} of M-theory or type-II string theories. 
It is also known that the symmetry is enhanced to $E_8$ or much 
larger (infinite-dimensional) 
upon reduction to three or lower dimensions 
\cite{MarcusSchwarz,NicolaiN=16,Julia1982,GebertNicolai}.

The $E$-series is a token of lower-dimensional M/typeIIA/typeIIB 
theories upon toroidal compactifications. The $D$-series, on the other hand, 
is known to appear as 
a similar symmetry group of the non-linear sigma model of the dimensionally 
reduced NS-NS sector supergravity, whose discrete subgroup is 
a T-duality of the toroidally compactified string theory \cite{MaharanaSchwarz}.
It is also very well known that the $A$-series is a symmetry 
of dimensionally reduced pure gravity \cite{Geroch,BreitenlohnerMaison}.
The $B$-series may be obtained as a reduction of the NS-NS sector 
coupled to an odd number of vector fields \cite{MaharanaSchwarz,Sen}, 
and $G_{2(+2)}$ has been shown to be the symmetry of the 
dimensionally reduced $D=5$ minimal
supergravity to three dimensions \cite{MizoguchiOhta}.
So what about the remaining simple Lie algebras?

As for $F_4$, many years ago it was anticipated that 
$F_{4(+4)}/(USp(6)\times SU(2))$ should be the sigma model of 
the dimensionally reduced $D=5$ {\em magical supergravity} of the 
simplest kind, reduced down to three dimensions \cite{GSTPhysLett}
\footnote{Among the $C$-series, which is also missing in the above description,  
$Sp(6,\RRsub)/U(3)$ $(Sp(6)=C_3)$ has also been shown to appear \cite{GSTNuclPhys} 
as a scalar coset of the same magical supergravity reduced to {\em four} dimensions. }
\footnote{See \cite{Karndumri1,Karndumri2} for the gaugings of the three-dimensional 
magical supergravities.}. 
Although the appearance of this particular quaternionic manifold has 
been justified on various grounds and is now believed to be true, 
a direct proof by performing a dimensional 
reduction of the supergravity and comparing to the 
construction to the coset group manifold seems to have never appeared 
in print. The aim of this letter is to fill this gap.

The direct proof of the $F_{4(+4)}/(USp(6)\times SU(2))$ coset structure 
has the following benefits:
\\
\noindent
(1) The direct dimensional reduction and the explicit construction of the 
coset sigma model enable us to find the precise relationship between 
the various components of the five-dimensional supergravity fields
and the relevant group elements. This allows us to use the $F_{4(+4)}$ 
global symmetry to generate a new supergravity solution from some known 
seed solution. Such a solution-generating technique utilizing the three- or 
four-dimensional global symmetry has been very powerful 
in deriving, for instance, the five-dimensional black hole solutions 
in five-dimensional minimal supergravity \cite{MizoguchiTomizawa}.
\\
\noindent
(2) By the above relationship between the supergravity fields and 
the group manifold one can also give group theoretical characterizations 
to some of the parameters and functions in the original magical 
supergravity Lagrangians. For example, as we show below, 
the $FFA$ coupling constants $C_{IJK}$ are identified as 
the structure constants of the commutation relations between generators 
both belonging to one of the ``Jordan pair'' in the decomposition 
\cite{FerraraMarraniZumino} of 
the quasi-conformal algebra of the relevant Jordan algebra. 
We will also find a Lie algebraic characterization of the functions 
of the scalars  
$\stackrel{\circ}a\!{}^{IJ}$ and $\stackrel{\circ}a\!{}_{IJ}$.

In fact, the procedure of the dimensional reduction itself is common to 
all the magical supergravity theories; the only difference is the 
range of the values of the indices of the vector and scalar fields.
Although the three-dimensional duality Lie algebras also allow a common 
decomposition in terms of the relevant Jordan algebras 
\cite{GSTPhysLett,GSTNuclPhys,GunaydinKoepsellNicolai,GunaydinPavlyk,
FerraraMarraniZumino}, 
in this letter  
we will work out in particular the $F_{4(+4)}$ case in detail. 
We expect, however, a similar identification or a characterization 
of the coupling constants 
and scalar metric functions may be done in other magical supergravities.

\section{Dimensional Reduction of $D=5$ Magical Supergravity}
The magical supergravities are $D=5$ $\mathcal{N}=2$ Einstein-Maxwell supergravities 
whose scalars of the vector multiplets constitute a coset sigma model with a symmetry 
group being a {\em simple} Lie group \cite{GSTNuclPhys}.
There exist four such theories, each of which is associated with 
one of the four division algebras 
$\mathbb{A}=\mathbb{R},\mathbb{C},\mathbb{H},\mathbb{O}$ 
and a rank-3 Jordan algebra ${\rm J}_3^{\AAsub}$ associated with it.
One of the characteristic features of these theories is that their five-dimensional 
Lagrangians as well as their dimensional reductions to four and three dimensions
universally contain scalar sigma models of the forms 
\cite{GSTNuclPhys,FerraraMarraniZumino}:
\beqa
\frac{\mbox{Str}_0({\rm J}_3^{\AAsub})}{\mbox{Aut}({\rm J}_3^{\AAsub})}~~(D=5),~~~
\frac{\mbox{M\"{o}}({\rm J}_3^{\AAsub})}{\widetilde{\mbox{Str}}_0({\rm J}_3^{\AAsub})
\times U(1)}~~(D=4),~~~
\frac{\mbox{qConf}({\rm J}_3^{\AAsub})}{\widetilde{\mbox{M\"{o}}}({\rm J}_3^{\AAsub})
\times SU(2)}~~(D=3),
\eeqa
where $\mbox{Aut}({\rm J}_3^{\AAsub})$, 
$\mbox{Str}_0({\rm J}_3^{\AAsub})$, 
$\mbox{M\"{o}}({\rm J}_3^{\AAsub})$
and
$\mbox{qConf}({\rm J}_3^{\AAsub})$
are respectively the automorphicm group, 
the reduced structure group, 
the superstructure group and 
the quasi-conformal group of the Jordan algebra  ${\rm J}_3^{\AAsub}$. 
$\widetilde{\phantom{AA}}$ denotes the corresponding compact form. 
There supergravity theories have been dubbed ``magical" \cite{GSTPhysLett} 
because these groups are precisely the elements of the 
``magic square" (see \cite{GSTPhysLett} and references therein), 
each Lie algebra ${\cal L}_{\AAsub,\AAsub'}$
of which allows  
the decomposition 
\beqa
{\cal L}_{\AAsub,\AAsub'}&=&{\rm D}_{\AAsub} \oplus {\rm D}_{{\rm J}_3^{\AAsub'}}
\oplus (\AA_0 \times ({\rm J}_3^{\AAsub'})_0),
\eeqa
where $\AA'=\RR,\CC,\HH$ and $\OO$ corresponds to 
$\mbox{Aut}$, $\mbox{Str}_0$, $\mbox{M\"{o}}$ and $\mbox{qConf}$, respectively. 
Here ${\rm D}_{\AAsub}$ and ${\rm D}_{{\rm J}_3^{\AAsub'}}$ are 
the generators of the automorphisms of $\AA$ and ${\rm J}_3^{\AAsub'}$, 
and $\AA_0$ and $({\rm J}_3^{\AAsub'})_0$ are the traceless generators.

The magical supergravity corresponding to the division algebra $\AA$
has $n=3(1+\mbox{dim}\mathbb{A})-1$ vector multiplets. 
Keeping only the bosonic terms, the Lagrangian is given by
\beqa
\label{67}
\mathcal{L}&=&
\frac{1}{2}E^{(5)}R^{(5)}
-\frac{1}{4}E^{(5)}{\stackrel{\circ}a\!{}}_{IJ}F^I_{MN}F^{JMN}
-\frac{1}{2}E^{(5)}s_{xy}(\partial_M\phi^x)(\partial^M\phi^y)\n
&&
+\frac{1}{6\sqrt{6}}C_{IJK}\epsilon^{MNPQR}F^I_{MN}F^J_{PQ}A^K_R,
\eeqa
where $E^{(5)}$ is the determinant of the f\"{u}nfbein,  
and $R^{(5)}$ is the scalar curvature in $D=5$.
${\stackrel{\circ}a\!{}}_{IJ}$ and $s_{xy}$ are functions of scalar fields $\phi^x$ 
which come from the vector multiplets 
and satisfy ${\stackrel{\circ}a\!{}}_{IJ}={\stackrel{\circ}a\!{}}_{JI}$ 
and $s_{xy}=s_{yx}$, respectively.
In particular, $s_{xy}$ is the metric of $n$-dimensional 
Riemannian space $\mathcal{M}$ which is parametrized 
by the scalar fields $\phi^x$, where $x,y,\dots$ take $1,2,\dots,n$.
$F^I_{MN}$ is the Maxwell field strength $2\partial_{[\mu}A_{\nu]}^I$.
$C_{IJK}$ is a constant and symmetric in all indices.
$M,N,\dots$ are the five-dimensional curved indices.
There are $n+1$ vector fields $A^I_\mu$ because the graviton multiplet has a 
single vector field, so that $I,J,\dots=1,2,\dots,n+1$.
\par
To reduce the dimensions to $D=3$, we set the f\"{u}nfbein and its inverse as 
\begin{equation}
\label{68}
{E^{(5)}}_M{}^A=\left(
\begin{array}{cc}
e^{-1}{E_\mu}^\alpha&B^m_\mu {e_m}^a\\
0&{e_m}^a
\end{array}
\right),\ \ \ \ \ 
{E^{(5)}}_A{}^M=\left(
\begin{array}{cc}
e{E_\alpha}^\mu&-e{E_\alpha}^\mu B_\mu^m\\
0&{e_a}^m
\end{array}
\right),
\end{equation}
where $A,B,\dots$ are the five-dimensional flat indices, 
$\mu,\nu,\dots$ and $\alpha,\beta,\dots$ are 
the three-dimensional curved and flat indices, 
$m,n,\dots$ and $a,b,\dots$ are 
the compact two-dimensional curved and flat indices, respectively.
Then we get the reduced Lagrangian
\begin{equation}
\label{69}
\begin{split}
\mathcal{L}~=~&\frac{1}{2}ER-\frac{1}{8}Ee^2g_{mn}B^m_{\mu\nu}B^{n\mu\nu}+\frac{1}{8}E\partial_\mu g^{mn}\partial^\mu g_{mn}-\frac{1}{2}E e^{-2}\partial_\mu e\partial^\mu e-\frac{1}{2}Es_{xy}(\partial_\mu\phi^x)(\partial^\mu\phi^y)\\
&-\frac{1}{2}E{\stackrel{\circ}a\!{}}_{IJ}g^{mn}\partial_\mu A^I_m\partial^\mu A^J_n-\frac{1}{4}Ee^2{\stackrel{\circ}a\!{}}_{IJ}F^{(3)I}_{\mu\nu}F^{(3)J\mu\nu}+\frac{1}{\sqrt{6}}C_{IJK}\epsilon^{\mu\nu\rho}\epsilon^{mn}F^I_{\mu\nu}\partial_\rho A^J_mA_n^K,
\end{split}
\end{equation}
where $B^m_{\mu\nu}=2\partial_{[\mu}B^m_{\nu]}$.
We define $F^{(3)I}_{\mu\nu}\equiv F'^I_{\mu\nu}+B^m_{\mu\nu}A_m^I$, where $F'^I_{\mu\nu}=2\partial_{[\mu}A'^I_{\nu]}$ is the field strength of the Kaluza-Klein invariant vector field $A'^I_\mu=A^I_\mu-B^m_\mu A^I_m$.\par
To dualize $A'^I_\mu$ and $B^m_\mu$ fields, we introduce Lagrange multipliers
\begin{align}
\mathcal{L}_{\text{Lag.mult.}}&=\epsilon^{\mu\nu\rho}\varphi_I\partial_\mu F'^I_{\nu\rho}+\frac{1}{2}\epsilon^{\mu\nu\rho}\psi_m\partial_\mu B^m_{\nu\rho}\nonumber\\
&\overset{\text{P.I.}}{=}-\epsilon^{\mu\nu\rho}F^{(3)I}_{\mu\nu}\partial_\rho\varphi_I-\frac{1}{2}\epsilon^{\mu\nu\rho}B^m_{\mu\nu}(\partial_\rho\psi_m+\partial_\rho A_m^I\varphi_I-A^I_m\partial_\mu\varphi_I).
\label{70}
\end{align}
Using the equations of motion for $F^{(3)I}_{\mu\nu}$ and $B^m_{\mu\nu}$, we obtain the 
dualized Lagrangian $\tilde{\mathcal L}\equiv\mathcal{L}+\mathcal{L}_{\text{Lag.mult.}}$:
\begin{align}
\tilde{\mathcal{L}}~=~&\frac{1}{2}ER+\frac{1}{8}E\partial_\mu g^{mn}\partial^\mu g_{mn}-\frac{1}{2}E e^{-2}\partial_\mu e\partial^\mu e-\frac{1}{2}Es_{xy}(\partial_\mu\phi^x)(\partial^\mu\phi^y)-\frac{1}{2}E{\stackrel{\circ}a\!{}}_{IJ}g^{mn}\partial_\mu A^I_m\partial^\mu A^J_n\nonumber\\
&-2Ee^{-2}{\stackrel{\circ}a\!{}}{}^{II'}\left(\frac{1}{\sqrt{6}}C_{IJK}\epsilon^{mn}\partial_\mu A_m^JA_n^K-\partial_\mu\varphi_I\right)\left(\frac{1}{\sqrt{6}}C_{I'J'K'}\epsilon^{m'n'}\partial^\mu A_{m'}^{J'}A_{n'}^{K'}-\partial^\mu\varphi_{I'}\right)\nonumber\\
&-Ee^{-2}g^{mn}\left(\frac{2}{3\sqrt{6}}C_{IJK}\epsilon^{pq}\partial_\mu A^I_pA^J_qA_m^K+\partial_\mu\psi_m+\partial_\mu A^I_m\varphi_I-A^I_m\partial_\mu\varphi_I\right)\nonumber\\
&\qquad\qquad\times\left(\frac{2}{3\sqrt{6}}C_{I'J'K'}\epsilon^{p'q'}\partial^\mu A^{I'}_{p'}A^{J'}_{q'}A^{K'}_n+\partial^\mu\psi_n+\partial^\mu A^{I'}_n\varphi_{I'}-A^{I'}_n\partial^\mu\varphi_{I'}\right).
\label{71}
\end{align}

\section{$F_{4(+4)}/(USp(6)\times SU(2))$ sigma model: the explicit proof}
In this section we prove that, if $\AA=\RR$ $(n=5)$, the sigma model part of the reduced 
Lagrangian (\ref{71}) constitutes 
the $F_{4(+4)}/(USp(6)\times SU(2))$ sigma model by an explicit construction.

The real form $F_{4(+4)}$
of the exceptional Lie algebra $F_4$ is decomposed into a sum of 
representations of the Lie algebra of a maximal subgroup $SL(3,\RR)\times SL(3,\RR)$
as
\beqa
{\bf 52}&=&({\bf 8},{\bf 1})\oplus({\bf 3},{\bf \bar 6})
\oplus({\bf \bar 3},{\bf 6})\oplus({\bf 1},{\bf 8}).
\eeqa
In spite of the notation, they are represented by real matrices.
Later we will identify the first $SL(3,\RR)$ as the global symmetry group 
arising from the reduction of the gravity sector from five to three 
dimensions, and the second one as the numerator group of the coset 
sigma-model scalars already 
existing in five dimensions. To distinguish them 
we call the first simply $SL(3,\RR)$ while the second $\widetilde{SL}(3,\RR)$.

Let $\hat E^i_{~j}$ ($i,j=1,2,3$) be generators of the $SL(3,\RR)$ algebra with a 
constraint $\hat E^1_{~1}+\hat E^2_{~2}+\hat E^3_{~3}=0$.
Similarly let $\hat{\tilde E}{}^{\tilde a}_{~\tilde b}$ ${\tilde a},{\tilde b}=1,2,3$ 
be generators of 
$\widetilde{SL}(3,\RR)$ with $\hat{\tilde E}{}^1_{~1}+\hat{\tilde E}{}^2_{~2}
+\hat{\tilde E}{}^3_{~3}=0$. Their commutations relations are
\beqa
{[}\hat E^i_{~j},~\hat E^{k}_{~l}
{]}&=&\delta^k_j \hat E^i_{~l} - \delta^i_l \hat E^k_{~j},
\label{F4com1}\\
{[}\hat{\tilde E}{}^{\tilde a}_{~\tilde b},\hat{\tilde E}{}^{\tilde c}_{~\tilde d}
{]}&=&\delta^{\tilde c}_{\tilde b} \hat {\tilde E}{}^{\tilde a}_{~\tilde d} 
- \delta^{\tilde a}_{\tilde d} \hat{\tilde  E}{}^{\tilde c}_{~\tilde b},\\
{[}\hat E^i_{~j},\hat{\tilde E}{}^{\tilde c}_{~\tilde d}
{]}&=&0.
\eeqa
We also introduce additional generators $E^I_i$, $E^{*i}_I$ 
($i=1,2,3$, $I=1,\ldots,6$) transforming respectively as
$({\bf 3},{\bf \bar 6})$, $({\bf \bar 3},{\bf 6})$ under $SL(3,\RR)\oplus \widetilde{SL}(3,\RR)$:
\beqa
{[}\hat E^i_{~j},E^{*k}_I
{]}&=&\delta^k_j E^{*i}_I,\\
{[}\hat E^i_{~j},E^I_{k}
{]}&=&-\delta^i_k E^I_j,\\
{[}\hat{\tilde E}{}^{\tilde a}_{~\tilde b},E^{*k}_I
{]}&=&\bar T^{\tilde a}_{~\tilde b}{}_I^{~J} E^{*i}_J,\\
{[}\hat{\tilde E}{}^{\tilde a}_{~\tilde b},E^I_{k}
{]}&=&T^{\tilde a}_{~\tilde b}{}^I_{~J}  E_i^J.
\eeqa
$\bar T^{\tilde a}_{~\tilde b}{}_I^{~J} $ and $T^{\tilde a}_{~\tilde b}{}^I_{~J} $ 
are respectively the ${\bf \bar 6}$ and 
${\bf  6}$ representation matrices of $\widetilde{SL}(3, \RR)$. 
In fact, in the present choice of the basis of the generators 
the structure constants satisfy
\beqa
\bar T^{\tilde a}_{~~\tilde b}{}_I^{~A}&=&
-T^{\tilde a}_{~~\tilde b}{}^A_{~\;I}.
\eeqa

Finally we 
set the commutation relations among two of these generators as
\beqa
{[}E^I_i,~E^{*j}_J
{]}&=&
-4\delta_J^I \hat E^j_{~i} 
+\delta^j_i D^{I\;~~\tilde b}_{~J\tilde a} \hat{\tilde E}{}^{\tilde a}_{~\tilde b},\\
{[}E^I_i,~E^J_j{]}&=&
+C^{IJK}\epsilon_{ijk}E^{*k}_K,\\
{[}E^{*i}_I,~E^{*j}_J{]}&=&
-C_{IJK}\epsilon^{ijk}E^K_{k},
\label{F4com11}
\eeqa
where $C_{IJK}=C^{IJK}$ are symmetric with respect to any permutation of indices, 
and 
\beqa
&&D^I_{~J}{}_{\tilde a}^{~\tilde b}=D^J_{~I}{}_{\tilde b}^{~\tilde a}.
\eeqa
Their actual values in the present basis are
\beqa
&&C^{123}=\sqrt{2},\n
&&C^{456}=+2,\n
&&C^{114}=C^{225}=C^{336}=-2,
\eeqa
and
\beqa
D^1_{~2}{}_2^{~1}=D^1_{~3}{}_1^{~3}&=&+2,\n
D^2_{~3}{}_3^{~2}&=&-2,\n
D^1_{~6}{}_3^{~2}=D^1_{~5}{}_2^{~3}&=&+2\sqrt{2},\n
D^3_{~4}{}_1^{~2}=D^3_{~5}{}_1^{~2}=D^2_{~4}{}_1^{~3}=D^2_{~6}{}_3^{~1}&=&-2\sqrt{2},\n
D^1_{~1}{}_1^{~1}=D^2_{~2}{}_2^{~2}=D^3_{~3}{}_3^{~3}&=&+2,\n
D^4_{~4}{}_1^{~1}=D^5_{~5}{}_2^{~2}=D^6_{~6}{}_3^{~3}&=&+4,
\eeqa
otherwise 0.
One may verify that the commutations relations (\ref{F4com1})-(\ref{F4com11}) 
close and generate the whole $F_{4(+4)}$ Lie algebra.

In fact, these commutations relations are derived from those among generators 
of a more tractable realization of $F_{4(+4)}$ in terms of the decomposition into 
representations of another maximal subalgebra $O(4,5)$:
\beqa
{\bf 52}&=&{\bf 36}\oplus{\bf 16},
\eeqa
where ${\bf 36}$ is the adjoint representation of $O(4,5)$ and 
${\bf 16}$ is the Majorana spinor representation. They are further decomposed 
into representations of $O(4,4)$ as  
\beqa
{\bf 52}&=&{\bf 28}\oplus{\bf 8_v}\oplus{\bf 8_s}\oplus{\bf 8_c},
\eeqa
which shows the hidden triality in $F_{4(+4)}$. The commutation relations 
among generators are
\beqa
{[}X^{ab},~X^{cd}
{]}&=&\eta^{bc}X^{ad}-\eta^{ac}X^{bd}-\eta^{bd}X^{ac}+\eta^{ad}X^{bc},\\
{[}X^{ab},~v^{c}
{]}&=&\eta^{bc}v^{a}-\eta^{ac}v^{b},\\
{[}v^{a},~v^{b}
{]}&=&-X^{ab},\\
{[}X^{ab},~s^{\alpha}
{]}&=&-\frac12 (\bar{\gamma}^{[a}\gamma^{b]})^\alpha_{~\beta}s^\beta,\\
{[}X^{ab},~c_{\alpha}
{]}&=&-\frac12 (\bar{\gamma}^{[a}\gamma^{b]})_\alpha^{~\beta}c_{\beta},\\
{[}v^{a},~s^{\alpha}
{]}&=&+\frac12 (\bar{\gamma}^{a})^{\alpha\beta}c_{\beta},\\
{[}v^{a},~c_{\alpha}
{]}&=&-\frac12 (\gamma^{a})_{\alpha\beta}s^{\beta},\\
{[}s^{\alpha},~s^{\beta}
{]}&=&-\frac12 (\bar{\gamma}_{a}\gamma_{b}C)^{\alpha\beta}X^{ab},\\
{[}c_{\alpha},~c_{\beta}
{]}&=&+\frac12 (\gamma_{a}\bar{\gamma}_{b}C)_{\alpha\beta}X^{ab},\\
{[}s^{\alpha},~c_{\beta}
{]}&=&+(\gamma_{a}C)_{\beta}^{~\alpha}v^{a},
\eeqa
where $X^{ab}=-X^{ba}\in {\bf 28}$ ($a,b=1,\ldots,8$), 
$v^{a}\in {\bf 8_v}$ ($a=1,\ldots,8$), 
$s^{\alpha}\in {\bf 8_s}$ ($\alpha=1,\ldots,8$), and
$c_{\alpha}\in {\bf 8_c}$ ($\alpha=1,\ldots,8$).
Here the conventions are 
$\eta^{ab}=\mbox{diag}(-1,-1,-1,-1,1,1,1,1)$, and
$\gamma^{a}$,
$\bar{\gamma}^{a}$ $(a=1,\ldots,8)$ 
are off-diagonal blocks of $O(4,4)$ gamma 
matrices in the Majorana-Weyl representation:
\beqa
\Gamma^a&=&\left(
\begin{array}{cc}& \bar{\gamma}^{a}\\
\gamma^{a}&
\end{array}
\right).
\eeqa
$C$ is the charge conjugation matrix satisfying
\beqa
C\gamma_{a}^T&=&-\gamma_{a}C,\\
C\bar{\gamma}_{a}^T&=&-\bar{\gamma}_{a}C.
\eeqa

The generators $\hat E^i_{~j}\in({\bf 8},{\bf 1})$ ($i,j=1,2,3$),
$\hat{\tilde E}{}^{\tilde a}_{~\tilde b}\in({\bf 1},{\bf 8})$ $(\tilde a,\tilde b=1,2,3)$, 
$E^I_i\in({\bf \bar 3},{\bf 6})$ and
$E^{*i}_I\in({\bf 3},{\bf \bar 6})$ ($i=1,2,3$, $I=1,\ldots,6$)
in the $SL(3,\RR)\times \widetilde{SL}(3,\RR)$ decomposition 
can be found as follows:
\begin{itemize}
\item{One can take $\hat E^i_{~j}$'s as the standard $SL(3,\RR)$ 
generators in the $O(3,3)$ subalgebra of $O(4,4)$.}
\item{In the remaining generators of $O(4,4)$ one can find three pairs of 
${\bf 3}$ and ${\bf\bar 3}$ of $SL(3,\RR)$.}
\item{Also in each of $v^a$, $s^\alpha$ and $c_\alpha$ one can find 
a single pair, in total another three pairs, of ${\bf 3}$ and ${\bf\bar 3}$ of 
$SL(3,\RR)$.}
\item{The remaining eight generators 
that do not belong to any of the above turn out to generate another $SL(3,\RR)$ 
algebra, $\widetilde{SL}(3,\RR)$.}
\item{Finally, one can verify that these six pairs of ${\bf 3}$ and ${\bf\bar 3}$ 
respectively transform 
as ${\bf\bar 6}$ and ${\bf 6}$  under $\widetilde{SL}(3,\RR)$.}
\end{itemize}

In terms of the $SL(3,\RR)\times\widetilde{SL}(3,\RR)$ decomposition, 
the whole $F_{4(+4)}$ generators are classified into ${\bf H}$ and ${\bf K}$, 
of which $F_{4(+4)}$ is a direct sum:
\beqa
F_{4(+4)}&=&{\bf H} \oplus {\bf K}.
\eeqa
${\bf H}$ consists of ``compact'' generators:
\beqa
{\bf H}&=&(\oplus_{i,j=1,2,3} \RR (\hat E^i_{~j}-\hat E^j_{~i}))
\oplus
(\oplus_{\tilde a,\tilde b=1,2,3} \RR (\hat{\tilde E}{}^{\tilde a}_{~\tilde b}-\hat{\tilde E}{}^{\tilde b}_{~\tilde a}))\n
&&
\oplus
(\oplus_{i=1,2,3;I=1,\ldots,6} \RR (E_i^{I}-E_I^{*i})).
\eeqa 
The Killing bilinear form on ${\bf H}$ is negative definite. It turns out that 
the independent $3+3+18=24$ generators of ${\bf H}$ generate 
$USp(6)\oplus SU(2)$.  The generators of this factorized $SU(2)$ are
\beqa
H_i&=&\frac12\left(
\hat E^{i+1}_{~~i+2}-\hat E^{i+2}_{~~i+1})
\right)
+\frac14
\left(
E^4_i-E^{*i}_4+E^5_i-E^{*i}_5+
E^6_i-E^{*i}_6
\right)
\eeqa
$(i=1,2,3)$, where the indices of $\hat E$ are defined modulo 3. 
$H_i$'s satisfy the $SU(2)$ commutation relations 
\beqa
{[}H_i,~H_j{]}&=&-2\epsilon_{ijk}H_k.
\eeqa

In fact, this $SU(2)$ is one of the irreducible $SU(2)$ subalgebra of 
$O(4)=SU(2)\oplus SU(2)$, which itself is an irreducible one of 
the maximal compact subalgebra $O(4)\oplus O(4)$ of $O(4,4)$.   
Thus they trivially commute with other compact generators 
contained in $O(4,5)=O(4,4)\oplus \oplus_{a=1,\ldots,8}\RR v_a$. It can also be verified that they also commute with compact generators made out of $s^\alpha$'s and $c_\alpha$'s.
The remaining orthogonal compliment in ${\bf H}$ consisting of 21 generators 
generates $USp(6)$.

On the other hand, ${\bf K}$ is spanned by all the ``noncompact'' generators:
\beqa
{\bf K}&=&(\oplus_{i,j=1,2,3} \RR (\hat E^i_{~j}+\hat E^j_{~i}))
\oplus
(\oplus_{\tilde a,\tilde b=1,2,3} \RR (\hat{\tilde E}{}^{\tilde a}_{~\tilde b}+\hat{\tilde E}{}^{\tilde b}_{~\tilde a}))\n
&&
\oplus
(\oplus_{i=1,2,3;I=1,\ldots,6} \RR (E_i^{I}+E_I^{*i})).
\eeqa 
The $52-28=24$ generators of ${\bf K}$ parametrize 
the ``physical'' degrees of freedom of the $F_{4(+4)}/(USp(6)
\times SU(2))$ nonlinear sigma model.

$F_{4(+4)}/(USp(6)
\times SU(2))$ is a symmetric space for which we denote the Cartan involution as $\tau$:
\beqa
{[}{\bf H},~{\bf H}{]}&\subset&{\bf H},\n
{[}{\bf K},~{\bf K}{]}&\subset&{\bf H},\\
{[}{\bf H},~{\bf K}{]}&\subset&{\bf K},
\eeqa 
\beqa
\tau({\bf H})=-{\bf H},~~~\tau({\bf K})=+{\bf K}.
\eeqa
As usual, to construct a coset nonlinear sigma model, we define 
some group element ${\cal V}$ and consider 
\beqa
{\cal M}&\equiv&\tau({\cal V}^{-1}){\cal V}.
\eeqa
Then the Lagrangian is given, up to a constant, by
\beqa
-\frac14 E^{(3)}\mbox{Tr}\partial_\mu{\cal  M}^{-1} \partial^\mu{\cal M}
&=&E^{(3)}\mbox{Tr}\left(\frac12\left(
\partial_\mu{\cal V}{\cal V}^{-1}+\tau\left(\partial_\mu{\cal V}{\cal V}^{-1}\right)
\right)\right)^2.
\label{TrdMdM-1}
\eeqa

In order to reproduce the dimensionally reduced Lagrangian (\ref{71}) of the 
magical supergravity, we take \footnote{Here we use 
dotted numbers for the flat local Lorentz (though Euclidean here) indices 
$a'=\dot 1,\dot 2$,  
to distinguish them from the curved tangent space indices $i'=1,2$ 
for the reduced dimensions. 
} 
\beqa
{\cal V}&=&{\cal V}_-{\cal V}_+,\\
{\cal V}_+&=&{\cal V}_+^{grav.}+{\cal V}_+^{scalar},\\
{\cal V}_+^{grav.}&=&
\exp\left(
\log e_{\dot 1}^{~1}\hat E^1_{~1}
+\log e_{\dot 2}^{~2}\hat E^2_{~2}
+\log e\;\hat E^3_{~3}
\right)\n
&&\cdot\exp\left(
-e_1^{~\dot 2} e_{\dot 2}^{~2} \hat E^1_{~2}\right)
\exp\left(
\psi_1\hat E^1_{~3}+\psi_2\hat E^2_{~3}
\right),
\\
 {\cal V}_+^{scalar}&=&\exp\left((\log{\bf\tilde  e}^{-1})_{\tilde a}^{~\tilde i}
 \hat{\tilde E}{}^{\tilde a}_{~\tilde i}\right)~~~~(\tilde a,\tilde i=1,2,3),
\eeqa
where we have taken the zweibein for the reduced dimensions to be
in the upper-triangular form
\beqa
e_{i'}^{~a'}&=&\left(
\begin{array}{cc} e_{1}^{~\dot 1}&e_{1}^{~\dot 2}\\
0&e_{2}^{~\dot 2}
\end{array}
\right)
\eeqa
so that 
\beqa
e&=&{\rm det}e_{i'}^{~a'}
=(e_{\dot 1}^{~1}e_{\dot 2}^{~2})^{-1},
\eeqa
and
\beqa
{\bf \tilde e}^{-1}&=&\left(
\begin{array}{ccc}
s_{11}&s_{12}&s_{13}~~\\
0&s_{22}&s_{23}~~\\
0&0&(s_{11}s_{22})^{-1}
\end{array}
\right)^{-1}.
\label{etilde}
\eeqa

For ${\cal V}_-$ we take 
\beqa
{\cal V}_-&=&\exp\left(
A_{i'}^I E_I^{*i'}
+\varphi_I E^I_3\right)
~~~(i'=1,2; ~ I=1,\ldots,6).
\eeqa
Then a straightforward calculation yields
\beqa
\partial_\mu {\cal V}{\cal V}^{-1}&=&
\partial_\mu {\cal V}_+{\cal V}_+^{-1}
+ {\cal V}_+(\partial_\mu {\cal V}_-{\cal V}_-^{-1}){\cal V}_+^{-1},\\
\partial_\mu {\cal V}_+{\cal V}_+^{-1}&=&
(e_{\dot 1}^{~1})^{-1}\partial_\mu e_{\dot 1}^{~1} \hat E^1_{~1}+
(e_{\dot 2}^{~2})^{-1}\partial_\mu e_{\dot 2}^{~2} \hat E^2_{~2}+
e^{-1}\partial_\mu e \hat E^3_{~3}\n
&&-e_{\dot 1}^{~1}(e_{\dot 2}^{~2})^{-1} \partial_\mu B\hat E^1_{~2}
+e^{-1}\left(
e_{\dot 1}^{~1}(\partial_\mu \psi_1-B\partial_\mu \psi_2)\hat E^1_{~3}
+e_{\dot 2}^{~2}\partial_\mu \psi_2 \hat E^2_{~3}
\right)\n
&&+\partial_\mu \tilde e_{\tilde a}^{~\tilde i}~\tilde e{}_{\tilde i}^{~\tilde b}
\hat{\tilde E}{}^{\tilde a}_{~\tilde b}
\label{dV+V+-1}
\eeqa
$\left(e_{a'}^{~i'}=\left(
\begin{array}{cr}
e_{\dot 1}^{~1}&-e_{\dot 1}^{~1} B\\
0&e_{\dot 2}^{~2}
\end{array}
\right)\right)$,
and
\beqa
 {\cal V}_+(\partial_\mu {\cal V}_-{\cal V}_-^{-1}){\cal V}_+^{-1}
&=&
e_{a'}^{~i'}\stackrel{\circ}f\!{}_I^{~A}\partial_\mu A_{i'}^I E_A^{*a'}\n
&&+e^{-1}\stackrel{\circ}f\!{}^I_{~A}\left(
\partial_\mu\varphi_I-\frac12 C_{JKI}\epsilon^{i'j'}A_{i'}^J\partial_\mu A_{j'}^K
\right)E^A_3\n
&&+e^{-1}e_{a'}^{~i'}
\left(
2 (A_{i'}^I\partial_\mu\varphi_I- \partial_\mu A_{i'}^I ~\varphi_I)
-\frac{2}3 C_{JKI} \epsilon^{j'k'}A_{i'}^I A_{j'}^J\partial_\mu A_{k'}^K 
\right)\hat E^{a'}_{~~3},\n
\label{V+dV-V--1V--1}
\eeqa 
where 
\beqa
\stackrel{\circ}f\!{}_I^{~A}&=&(\exp((\log \tilde {\bf e}^{-1})_{\tilde a}^{~\tilde b}
\bar T^{\tilde a}_{~~\tilde b}))_I^{~A},\\
\stackrel{\circ}f\!{}^I_{~A}&=&(\exp((\log \tilde {\bf e}^{-1})_{\tilde a}^{~\tilde b}
T^{\tilde a}_{~~\tilde b}))^I_{~A}
\eeqa
are respectively the ${\bf \bar 6}$ and ${\bf 6}$ representation matrices 
of the $\widetilde{SL}(3,\RR)$ group element $\tilde {\bf e}^{-1}$ 
(\ref{etilde}).

Plugging (\ref{dV+V+-1})(\ref{V+dV-V--1V--1}) 
into (\ref{TrdMdM-1}), $\frac12(\partial_\mu{\cal V}{\cal V}^{-1}
+\tau(\partial_\mu{\cal V}{\cal V}^{-1}))$ projects out the ${\bf H}$ piece of 
$\partial_\mu{\cal V}{\cal V}^{-1}$, leaving only the ${\bf K}$ piece.
This amounts to the replacements 
\beqa
\hat E^i_{~j}&\longrightarrow&\frac12(\hat E^i_{~j}+\hat E^j_{~i}),\n
\hat{\tilde E}{}^{\tilde a}_{~\tilde b}&\longrightarrow&\frac12(\hat{\tilde E}{}^{\tilde a}_{~\tilde b}
+\hat{\tilde E}{}^{\tilde b}_{~\tilde a}),\n
E^I_i&\longrightarrow&\frac12(E^I_i+E_I^{*i}),\n
E_I^{*i}&\longrightarrow&\frac12(E^I_i+E_I^{*i})
\eeqa
in $\partial_\mu{\cal V}{\cal V}^{-1}$.
Thus, using the invariant bilinear form computed in the adjoint representation 
normalized by twice the dual Coxeter number $2 h^\vee_{F_4}=18$:
\footnote{It is simpler to use $E^a_{~b}$, $\tilde E^{\tilde a}_{~\tilde b}$ than to use 
hatted generators to compute traces, where 
$\hat E^a_{~b}=E^a_{~b}-\frac13\delta^a_b(E^1_{~1}+E^2_{~2}+E^3_{~3})$ 
and similarly for $\hat{\tilde E}{}^{\tilde a}_{~\tilde b}$}
\beqa
\frac1{18}
{\rm Tr}E^a_{~b}E^c_{~d}&=&\delta^c_b \delta^a_d~~~(a,b,c,d=1,2,3),
\n
\frac1{18}
{\rm Tr}\tilde E^{\tilde a}_{~\tilde b}\tilde E^{\tilde c}_{~\tilde d}
&=&2\delta^{\tilde c}_{\tilde b} \delta^{\tilde a}_{\tilde d}
~~~(\tilde a,\tilde b,\tilde c,\tilde d=1,2,3),
\n
\frac1{18}{\rm Tr}E^A_{a}E_B^{*b}&=&
4\delta^b_a \delta^A_B~~~(a,b=1,2,3;~~A,B=1,\ldots,6) ,
\n
\mbox{otherwise}&=&0,
\eeqa
we obtain
\beqa
\frac1{72}{\rm Tr}\partial_\mu{\cal M}^{-1}
\partial^\mu {\cal M}
&=&\frac14 \partial_\mu g^{ij} \partial_\mu g_{ij} 
-e^{-2}\partial_\mu e \partial^\mu e
+\frac12 \partial_\mu \tilde g^{\tilde i\tilde j} \partial_\mu \tilde g_{\tilde i\tilde j} 
-2 g^{ij}\!\stackrel{\circ}a\!{}_{IJ} \partial_\mu A_i^I
 \partial^\mu A_j^J\n
&&-2 e^{-2}
\stackrel{\circ}a\!{}^{IJ}
\left(
\partial_\mu\varphi_I
-\frac12 C_{KLI}\epsilon^{kl}A^K_k \partial_\mu A^L_l
\right)\n
&&
~~~~~~~~~~~~\cdot
\left(
\partial^\mu\varphi_J
-\frac12 C_{K'L'J}\epsilon^{k'l'}A^K_{k'} \partial_\mu A^L_{l'}
\right)
\n
&&-\frac12 e^{-2} g^{ij} 
\left(
\partial_\mu\psi_i -2(\varphi_I\partial_\mu A^I_i-\partial_\mu\varphi\; A^I_i)
-\frac23 C_{KLI}\epsilon^{kl} A^K_k \partial_\mu A^L_l \; A^I_i
\right)\n
&&~~~~~~~\cdot \left(
\partial^\mu\psi_j -2(\varphi_I\partial_\mu A^J_j-\partial_\mu\varphi\; A^J_j)
-\frac23 C_{K'L'J}\epsilon^{k'l'} A^{K'}_{k'} \partial_\mu A^{L'}_{l'}A^J_j
\right).\n
&&
\eeqa
This final form of the sigma model coincides with $2E^{-1}$ times the 
dimensionally reduced Lagrangian (\ref{71}) obtained in the previous section 
with the rescalings  
\beqa
A^I_i\rightarrow \frac{A^I_i}{\sqrt 2},~~\varphi_I\rightarrow \frac{\varphi_I}{\sqrt 2},
~~~\psi_i\rightarrow 2\psi_i,~~~C_{IJK}\rightarrow\frac4{\sqrt 3}C_{IJK}. 
\eeqa
This complete the direct proof of the equivalence of the dimensionally reduced Lagrangian 
of the magical supergravity to the $F_{4(+4)}/(USp(6))\times SU(2))$ 
nonlinear sigma model.

\section{Conclusions and Discussion: Other magical supergravities}
In this letter we have shown the direct relationship between the (bosonic part of the) 
simplest of the four magical theories reduced to three dimensions and 
the $F_{4(+4)}/(USp(6))\times SU(2))$ coset sigma model.
As we mentioned in Introduction, these 
relations will be used to generate various new supergravity solutions by 
applying $F_{4(+4)}$ transformations to some known solutions of 
this magical supergravity. 

We can give some Lie algebraic characterizations to various geometrical 
quantities defined in the supergravity Lagrangian: 
\begin{itemize}
\item{$C_{IJK}$'s are the structure constants of the commutation relations between 
generators both belonging to $({\bf 3},{\bf \bar 6})$. In particular $I=1,\ldots,6$ 
are the indices for a symmetric tensor representation ${\bf \bar 6}$ of the
$SL(3,\RR)$, which is the numerator group of the scalar coset $SL(3,\RR)/SO(3)$
already existing 
in five dimensions.}
\item{$\stackrel{\circ}a\!{}^{IJ}$ and $\stackrel{\circ}a\!{}_{IJ}$ are nothing but 
the ${\bf 6}$ and ${\bf \bar 6}$ representation matrices of
the metric of the reduced two dimensions viewed as an $SL(3,\RR)$ group element. }
\end{itemize}

We note that the structures we found here are very similar to the dimensionally
reduced  
eleven-dimensional supergravity or the $D=5$ minimal supergravity to 
three dimensions \cite{MizoguchiE10,MizoguchiOhta,MizoguchiSchroder}, 
whose sigma models are respectively
$E_{8(+8)}/SO(16)$ and $G_{2(+2)}/SO(4)$.

In all the magical supergravity theories, the number of the original scalars ($=n$) 
is always one less than the number of the abelian gauge fields. 
In the simplest magical case considered in this letter, this is the number of 
the dimension of the {\em symmetric tensor} representation, which is 6.
In fact, for the other three magical cases, we can also find representations 
of the numerator group of the coset whose dimensions are {\em pricisely} 
one more than the dimensions of the coset of the respective theories 
\cite{FerraraMarraniZumino}:

\begin{itemize}
\item{ ${\rm J}_3^{\CCsub}$ magical}:
\beqa
E_{6(+2)}&\supset& SL(3,\RR)\times SL(3,\CC)=SL(3,\RR)\times (SL(3,\RR)\times SL(3,\RR))\\
{\bf 78}&=&({\bf 8},({\bf 1},{\bf 1}))\oplus({\bf 3},({\bf \bar 3},{\bf \bar 3}))
\oplus({\bf \bar 3},({\bf 3},{\bf 3}))\oplus({\bf 1},({\bf 8},{\bf 1}))\oplus({\bf 1},({\bf 1},{\bf 8})).
\eeqa
The dimension of the five-dimensional scalar coset is
\beqa
{\rm dim}
\frac{SL(3,\CC)}{SU(3)}&=&8,
\eeqa
so the index $I$ runs from $1$ to $9$. This agrees with the fact that 
the {\em direct product} representation $({\bf 3},{\bf 3})$ or $({\bf \bar 3},{\bf \bar 3})$ 
is nine-dimensional.
\item{ ${\rm J}_3^{\HHsub}$ magical}:
\beqa
E_{7(-5)}&\supset& SL(3,\RR)\times SU^*(6)\\
{\bf 133}&=&({\bf 8},{\bf 1})\oplus({\bf 3},{\bf \bar 15})
\oplus({\bf \bar 3},{\bf 15})\oplus({\bf 1},{\bf 35}).
\eeqa
The dimension of the coset is
\beqa
{\rm dim}
\frac{SU^*(6)}{USp(6)}&=&14.
\eeqa
In this case the relevant representations are the {\em rank-2 antisymmetric tensor} 
representations, which are ${\bf 15}$ and ${\bf \overline{15}}$.
\item{ ${\rm J}_3^{\OOsub}$ magical}:
\beqa
E_{8(-24)}&\supset& SL(3,\RR)\times E_{6(-26)}\\
{\bf 133}&=&({\bf 8},{\bf 1})\oplus({\bf 3},{\bf \overline{27}})
\oplus({\bf \bar 3},{\bf 27})\oplus({\bf 1},{\bf 78}).
\eeqa
In this case
\beqa
{\rm dim}
\frac{E_{6(-26)}}{F_4}&=&26.
\eeqa

This also agrees with the existence of the {\em fundamental} 
${\bf 27}$ and ${\bf \overline{27}}$ representations of $E_6$ with 
the above decomposition of $E_{8(-24)}$.
\end{itemize}

In view of this common structure of decompositions (known as 
the decomposition of the quasi-conformal algebra of the Jordan 
algebra in terms of the super-Ehlers' algebra \cite{FerraraMarraniZumino}
), we expect the same characterization for $C_{IJK}$ or 
$\stackrel{\circ}a\!{}^{IJ}$ and $\stackrel{\circ}a\!{}_{IJ}$ will be possible 
for the other three magical supergravity theories. 
To show this the realizations worked out in \cite{GunaydinKoepsellNicolai} 
will be useful. Work along this line is in progress.

\section*{Acknowledgments} 
We would like to thank A. Ishibashi, H. Kodama and  S. Tomizawa 
for discussions. A conversation had with H. Nicolai some time ago 
has been also useful, for which he is also acknowledged. 
The work of S.~M. is supported by 
Grant-in-Aid
for Scientific Research  
(C) \#25400285, 
(C) \#16K05337 
and 
(A) \#26247042
from
The Ministry of Education, Culture, Sports, Science
and Technology of Japan.


\begin{thebibliography}{99}


\bibitem{CJ}
  E.~Cremmer and B.~Julia,
  Phys.\ Lett.\ B {\bf 80} (1978) 48;
  Nucl.\ Phys.\ B {\bf 159} (1979) 141.

\bibitem{CJS}
E.~Cremmer, B.~Julia and J.~Scherk,
  Phys.\ Lett.\ B {\bf 76} (1978) 409.

\bibitem{JuliaGroupDisintegrations}
B.~Julia,
  Conf.\ Proc.\ C {\bf 8006162} (1980) 331.

\bibitem{Keurentyes}
A.~Keurentjes,
  Nucl.\ Phys.\ B {\bf 658} (2003) 303
  [hep-th/0210178].
  
\bibitem{HullTownsend}
C.~M.~Hull and P.~K.~Townsend,
  Nucl.\ Phys.\ B {\bf 438} (1995) 109
  [hep-th/9410167].

\bibitem{MarcusSchwarz}
N.~Marcus and J.~H.~Schwarz,
  Nucl.\ Phys.\ B {\bf 228} (1983) 145.
  
\bibitem{NicolaiN=16}
H.~Nicolai,
  Phys.\ Lett.\ B {\bf 194} (1987) 402.

\bibitem{Julia1982}
B. Julia, Kac-Moody Symmetry of Gravitation and Supergravity Theories,
AMS-SIAM Summer Seminar on Applications of 
Group Theory in Physics and Mathematics, Chicago (1982).


\bibitem{GebertNicolai}
R. W. Gebert and H. Nicolai, E10 for Beginners, Gursey Memorial Conference I: On Strings and Symmetries, Istanbul (1994).


\bibitem{Geroch}R. Geroch, J. Math. Phys. 13 (1972) 394.

\bibitem{BreitenlohnerMaison}
P. Breitenlohner and D. Maison, Ann. Inst. Henri Poincar\'{e}, 46 (1987) 216.

 \bibitem{MaharanaSchwarz}
 J.~Maharana and J.~H.~Schwarz,
  Nucl.\ Phys.\ B {\bf 390} (1993) 3
  [hep-th/9207016].
 
 \bibitem{Sen} 
 A.~Sen,
  Int.\ J.\ Mod.\ Phys.\ A {\bf 9} (1994) 3707
  [hep-th/9402002].
 
 
\bibitem{MizoguchiOhta}
  S.~Mizoguchi and N.~Ohta,
  Phys.\ Lett.\ B {\bf 441} (1998) 123
  [hep-th/9807111].


\bibitem{GSTNuclPhys} 
M.~Gunaydin, G.~Sierra and P.~K.~Townsend,
  Nucl.\ Phys.\ B {\bf 242} (1984) 244.


\bibitem{GSTPhysLett}
  M.~Gunaydin, G.~Sierra and P.~K.~Townsend,
  Phys.\ Lett.\ B {\bf 133} (1983) 72.

\bibitem{Karndumri1}
  P.~Karndumri,
  JHEP {\bf 1208} (2012) 007
  doi:10.1007/JHEP08(2012)007
  [arXiv:1206.2150 [hep-th]].

\bibitem{Karndumri2}
  P.~Karndumri,
  JHEP {\bf 1512} (2015) 153
  doi:10.1007/JHEP12(2015)153
  [arXiv:1509.07431 [hep-th]].

\bibitem{MizoguchiTomizawa}
S.~Mizoguchi and S.~Tomizawa,
  Phys.\ Rev.\ D {\bf 84} (2011) 104009
  [arXiv:1106.3165 [hep-th]].\\
S.~Tomizawa and S.~Mizoguchi,
  Phys.\ Rev.\ D {\bf 87} (2013) no.2,  024027
  [arXiv:1210.6723 [hep-th]].
  
  

\bibitem{GunaydinKoepsellNicolai}
  M.~Gunaydin, K.~Koepsell and H.~Nicolai,
  Commun.\ Math.\ Phys.\  {\bf 221} (2001) 57
  [hep-th/0008063].

\bibitem{GunaydinPavlyk}
  M.~Gunaydin and O.~Pavlyk,
  JHEP {\bf 0501} (2005) 019
  [hep-th/0409272]
  
\bibitem{FerraraMarraniZumino}
S.~Ferrara, A.~Marrani and B.~Zumino,
  J.\ Phys.\ A {\bf 46} (2013) 065402
  [arXiv:1208.0347 [math-ph]].



\bibitem{MizoguchiE10}
  S.~Mizoguchi,
  Nucl.\ Phys.\ B {\bf 528} (1998) 238
  [hep-th/9703160].

\bibitem{MizoguchiSchroder}
  S.~Mizoguchi and G.~Schroder,
  Class.\ Quant.\ Grav.\  {\bf 17} (2000) 835
  [hep-th/9909150].

\end{thebibliography}
\end{document}